\documentclass[aps,prl,twocolumn,showpacs,preprintnumbers,amsmath,amssymb, superscriptaddress]{revtex4}

\usepackage{graphicx}
\usepackage{dcolumn}
\usepackage{bm}
\usepackage{color, float}

\newcommand{\ket}[1]{\left | #1 \right \rangle}
\newcommand{\bra}[1]{\left \langle #1 \right |}

\usepackage{amssymb}
\usepackage{amsmath}
\def\R{{\mathbb R}}
\def\C{{\mathbb C}}

\def\H{{\cal H}}
\def\S{{\cal S}}
\def\X{{\cal X}}
\def\d{{\delta}}
\def\m{{\mu}}
\def\r{{\rho}}
\def\th{{\theta}}
\def\Th{{\Theta}}
\def\Tr{{\rm Tr\,}}

\begin{document}

\preprint{APS/123-QED}

\title
{
Experimental Demonstration of Adaptive Quantum State Estimation
}

\author{Ryo Okamoto}
\affiliation{%
Research Institute for Electronic Science, Hokkaido University, Kita-ku
Sapporo 001--0020, Japan
}%
\affiliation{%
The Institute of Scientific and Industrial Research, Osaka University,
Mihogaoka 8-1, Ibaraki, Osaka 567-0047, Japan
}%

\author{Minako Iefuji}
\affiliation{%
Research Institute for Electronic Science, Hokkaido University, Kita-ku
Sapporo 001--0020, Japan
}%
\affiliation{%
The Institute of Scientific and Industrial Research, Osaka University,
Mihogaoka 8-1, Ibaraki, Osaka 567-0047, Japan
}%

\author{Satoshi Oyama}
\affiliation{%
Research Institute for Electronic Science, Hokkaido University, Kita-ku
Sapporo 001--0020, Japan
}%
\affiliation{%
The Institute of Scientific and Industrial Research, Osaka University,
Mihogaoka 8-1, Ibaraki, Osaka 567-0047, Japan
}%

\author{Koichi Yamagata}
\affiliation{%
Department of Mathematics, Osaka University,
1-1 Machikaneyama, Toyonaka 560-0043, Osaka, Japan
}%

\author{Hiroshi Imai}
\affiliation{%
University of Pavia, Dipartimento di Fisica A. Volta
Via Bassi 6, 27100 Pavia, Italy
}%

\author{Akio Fujiwara}
\email{fujiwara@math.sci.osaka-u.ac.jp}
\affiliation{%
Department of Mathematics, Osaka University,
1-1 Machikaneyama, Toyonaka 560-0043, Osaka, Japan
}%

\author{Shigeki Takeuchi}
\email{takeuchi@es.hokudai.ac.jp}
\affiliation{%
Research Institute for Electronic Science, Hokkaido University, Kita-ku
Sapporo 001--0020, Japan
}%
\affiliation{%
The Institute of Scientific and Industrial Research, Osaka University,
Mihogaoka 8-1, Ibaraki, Osaka 567-0047, Japan
}%

\date{\today}
%

\begin{abstract}
The first experimental demonstration of an adaptive quantum state estimation (AQSE) is reported. The strong consistency and asymptotic efficiency of AQSE have been mathematically proven [J. Phys. A:Math. Gen. 39 12489 (2006)]. In this Letter, the angle of linear polarization of single photons, or the phase parameter between the right and the left circularly polarization, is estimated using AQSE, and the strong consistency and asymptotic efficiency are experimentally verified. AQSE will provide a general useful method in both quantum information processing and metrology.
\end{abstract}

\maketitle
%

Quantum theory is inherently statistical. This entails repetition of experiments over a number of identically prepared quantum objects, for example, quantum states, if one wants to know the ``true state'' or the ``true value'' of the parameter that specifies the quantum state  \cite{James-2001, Matteo-2009, Bisio-2009, Hentschel-2011}.
Such an estimation procedure is particularly important for quantum communication and quantum computation \cite{Niel-2000},
and is also indispensable to quantum metrology \cite{Giovannetti2004, Na-2007, Okamoto-2008, Xiang-2010, Jones-2009}. 
In applications, one needs to design the estimation procedure in such a way that the estimated value of the parameter should be close to the true value (consistency), and that the uncertainty of the estimated value should be as small as possible (efficiency) for a given limited number of samples.
In order to realize these requirements, Nagaoka advocated an adaptive quantum state estimation (AQSE) procedure \cite{Na-1988,Na-1989}, and recently Fujiwara proved the strong consistency and asymptotic efficiency for AQSE \cite{{Fuj-2006},{Fuj-2011}}.

In this letter, we report the first experimental demonstration of AQSE using photons. The angle of a half wave plate (HWP) that initializes the linear polarization of input photons is estimated using AQSE. A sequence of AQSE is carried out with 300 input photons, and the sequence is repeated 500 times for four different settings of HWP. The statistical analysis of these results verifies the strong consistency and asymptotic efficiency of AQSE.  Recently, it has been mathematically proven that the precision of AQSE outperforms the conventional state tomography \cite{Yamagata}.
It is thus expected that AQSE will provide a useful methodology in the broad area of quantum information processing, communication, and metrology.

%

Let us first explain AQSE in detail. 
For simplicity, we restrict ourselves to one-dimensional {\em quantum statistical model} $\S=\{\r_\th;\,\th\in\Th\,(\subset\R)\}$, a smooth parametric family of density operators on a Hilbert space $\H$ having a one-dimensional parameter $\th$.
Our aim is to estimate the true value of $\th$ by means of a certain quantum estimation scheme. 
An {\em estimator} is represented by a pair $(M,\check\th)$, where $M=\{M(x);\, x\in\X\}$ is a positive operator-valued measure (POVM) that takes values on a set $\X$, and $\check\th:\X\to\Th$ is a map that gives the estimated value $\check\th(x)$ from each observed data $x\in\X$. The observed data $x\in\X$ has probability density 
\begin{equation}\label{eqn:density}
 f(x;\th,M):=\Tr\r_\th M(x),
\end{equation}
which depends on both the parameter $\th$ and the measurement $M$.

In traditional statistics, it is often the case to confine our attention to unbiased estimators. 
An estimator  $(M,\check\th)$ is called {\em unbiased} if 
\begin{equation}\label{eqn:UE}
  E_{\th}[M,\check\th]=\th
\end{equation}
is satisfied for all $\th\in\Th$,  where $E_{\th}[\;\cdot\;]$ denotes the expectation with respect to the density (\ref{eqn:density}). 
It is well known \cite{Helstrom:1976} that an unbiased estimator $(M,\check\th)$ satisfies the quantum Cram\'er-Rao inequality 
$V_{\th}[M,\check\th]\ge \left(J_{\th}\right)^{-1}$, 
where $V_{\th}[\;\cdot\;]$ denotes the variance, and $J_\th$ is the quantum Fisher information of the model $\S$ defined by
$J_\th:=\Tr \r_\th L_{\th}^2$, 
where $L_{\th}$ is the symmetric logarithmic derivative (SLD) defined by the self-adjoint operator satisfying the equation $\frac{d\r_\th}{d\th}=\frac{1}{2}\left(L_{\th}\r_\th+\r_\th L_{\th}\right).$

In quantum statistics, however, it is regarded that unbiasedness is too restrictive a requirement, and we usually weaken the condition to a ``local'' one. 
An estimator $(M,\check\th)$ is called {\em locally unbiased} \cite{Holevo:1982} at a given point 
$\th_0\in\Theta$ if the condition (\ref{eqn:UE}) is satisfied around $\th=\th_0$ up to the first order of the Taylor expansion, that is, if $E_{\th_0}[M,\check\th]=\th_0$ and 
$\left.\frac{d}{d\th} E_\th[M,\check\th]\right|_{\th=\th_0}=1$ hold.
Clearly, an estimator is unbiased if and only if it is locally unbiased at all $\th\in\Th$. 
A crucial observation is that an estimator $(M,\check\th)$ that is locally unbiased at $\th_0$ also satisfies the quantum Cram\'er-Rao inequality
\begin{equation}\label{eqn:Cramer-Rao}
 V_{\th_0}[M,\check\th]\ge \left(J_{\th_0}\right)^{-1}
\end{equation}
at $\th=\th_0$, and that the lower bound in (\ref{eqn:Cramer-Rao}) is achievable for any one-dimensional quantum statistical model $\S$. To put it differently, the best locally unbiased estimator (LUE) for the parameter $\th$ at $\th=\th_0$ is the one that satisfies $V_{\th_0}[M,\check\th]=\left(J_{\th_0}\right)^{-1}$. 
\begin{figure}
\begin{center}
\includegraphics*[width=6cm]{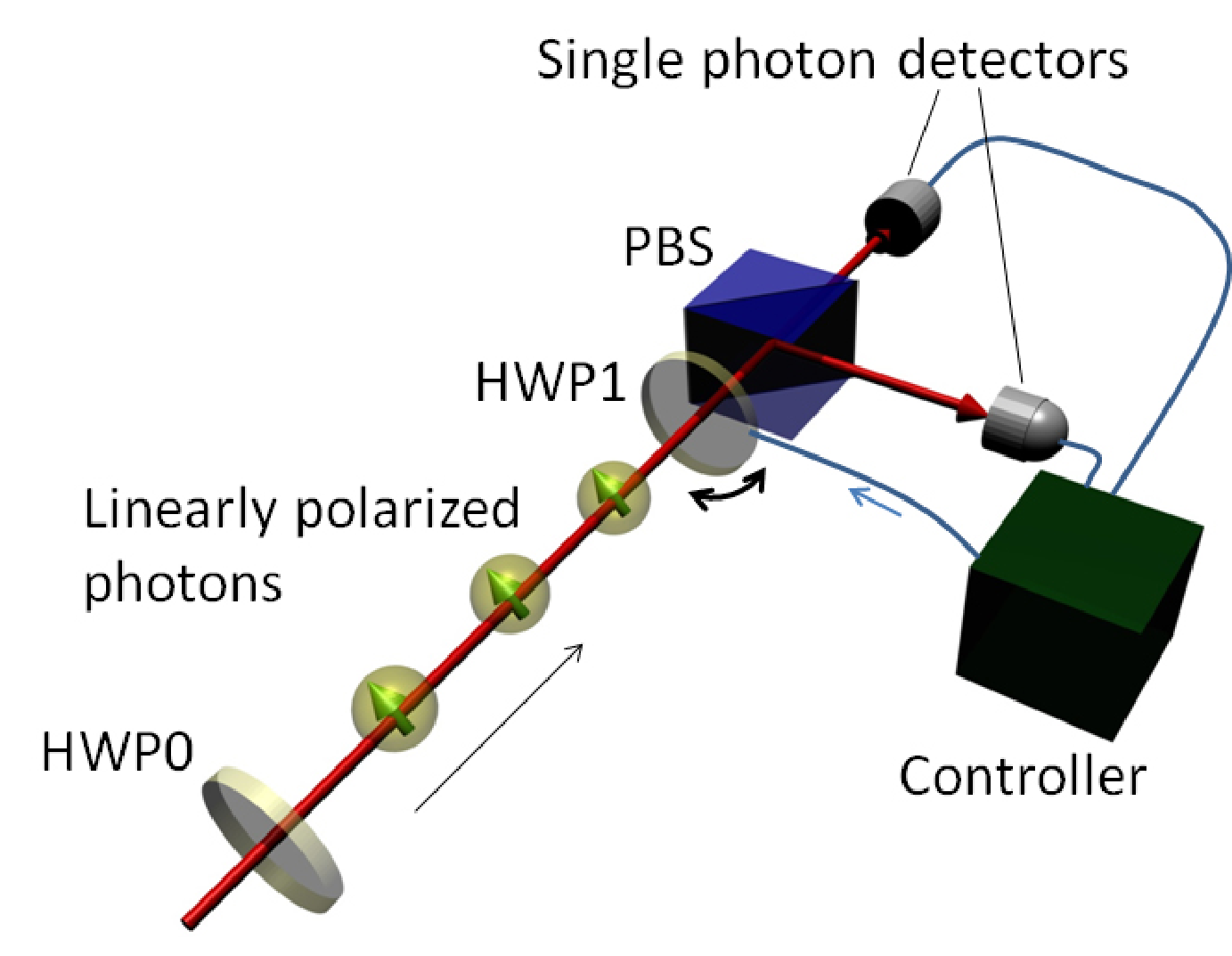}
\caption{Schematic of adaptive quantum state estimation. Photons are linearly polarized with a polarization direction determined by HWP0. The polarization is analyzed by HWP1 and the polarizing beam splitter (PBS). The controller sets HWP1 to an angle calculated on the basis of the photon measurement results.}
\end{center}
\label{schem1}
\end{figure}

Here we encounter a difficulty which often becomes the target of criticism: since the best LUE for estimating the parameter $\th$ depends, in general, on the unknown parameter $\th$ itself, the estimation strategy based on LUEs would be infeasible. 
In a different yet analogous context, Cochran \cite{Cochran} ingeniously described this kind of dilemma as follows: 
``You tell me the value of $\th$ and I promise to design the best experiment for estimating $\th$.''  

To surmount this difficulty, Nagaoka \cite{Na-1988,Na-1989} advocated an adaptive quantum state estimation (AQSE) scheme as follows. Suppose that, by prior investigation of the quantum statistical model $\S$, one has the list of optimal LUEs $\left(M(\;\cdot\;;\th), \check\th(\;\cdot\;;\th)\right)$ for each $\th\in\Th$. 
One begins with an arbitrary initial guess $\hat\th_0\in\Th$, and applies the measurement $M(\;\cdot\;;\hat\th_0)$ that is optimal at $\hat\th_0$. Suppose the data $x_1$ is observed, one then applies the maximum likelihood method to the likelihood function $L_1(\th)=f(x_1;\th, M(\;\cdot\;;\hat\th_0))$, to obtain the next guess $\hat\th_1$. At stage $n$ ($\ge 2$), one applies the measurement $M(\;\cdot\;;\hat\th_{n-1})$, where $\hat\th_{n-1}$ is the maximum likelihood estimator (MLE) obtained at the previous stage. The likelihood function is then given by $L_n(\th):=\prod_{i=1}^n\; f(x_i; \th, M(\;\cdot\;;\hat\th_{i-1})),$
where $x_i$ is the observed data at stage $i$, and one obtains the $n$th MLE $\hat\th_n$ that maximizes $L_n(\th)$. 
It is quite natural to expect that the sequence $\hat\th_n$ of MLEs would converge to the true value of the parameter $\th$. 
In fact, under certain regularity conditions, it can be shown that the sequence $\hat\th_n$  is strongly consistent and asymptotically efficient \cite{{Fuj-2006},{Fuj-2011}}. 

%
Now let us discuss the implementation of AQSE using photons (Fig. 1). Here the unknown parameter is the angle $\theta$ of HWP0, which determines the phase $\phi$ between right and left circularly polarizations of input photons by the relation $\phi = 4\theta$.  An arbitrary linear polarization can be described using right and left circular polarizations as follows:
\begin{equation}
\label{eqn:model}
\ket{\psi}
= \frac{1}{\sqrt{2}}(\ket{R}+e^{i \phi} \ket{L})
= \cos(\frac{\phi}{2})\ket{H}+\sin(\frac{\phi}{2})\ket{V}.
\end{equation}
By changing the angle of the half wave plate (HWP1), we can adjust the measurement basis.
For such measurement, the POVM having optimal estimation capability is given by
\begin{equation}\label{eqn:POVM}
M(\theta)
=(M(1;\theta), M(2;\theta))
=\left( \ket{\xi} \bra{\xi}, I-\ket{\xi} \bra{\xi} \right),
\end{equation}
where $\bra{\xi}=\left( \cos\left(2\theta+\frac{\pi}{4}\right),\, \sin\left(2\theta+\frac{\pi}{4}\right) \right).$ By applying the POVM $M(\theta)$ to the input state $\ket{\psi(\theta)}:=\ket{\psi}$, one obtains the probability distribution on $\X:=\{1,2\}$ which is isomorphic to the fair coin flipping.


The drawback to realizing this measurement is that the optimal POVM $M(\theta)$ depends on the unknown value of the parameter $\theta$ \cite{Comment1}.
We can avoid this drawback by adopting an AQSE as follows.
We begin by setting the initial log-likelihood function to be $l_{0}(\theta)=0$, 
and then start inputting and detecting photons one by one. 
For $n$th photon, we apply the measurement $M(\hat{\theta}_{n-1})$ which depends on the latest MLE $\hat{\theta}_{n-1}$. Let $x_n \in\X$ be the outcome indicating which detector has been lit. 
The log-likelihood function is then updated by the formula
\begin{equation}\label{eqn:elln}
l_{n}(\theta):=l_{n-1}(\theta)+\log \bra{\psi(\theta)}M(x_n;\hat{\theta}_{n-1}) \ket{\psi(\theta)},
\end{equation}
and the $n$th MLE is given by $\hat{\theta}_{n}=\arg\max_\theta l_{n}(\theta).$ Let us denote the true value of the parameter $\theta$ by $\theta^t$. It is known \cite{Fuj-2006, Fuj-2011} that the sequence $\hat\theta_n$ of MLEs converges to the true value $\theta^t$ with probability one (strong consistency) and that the distributions of the random variables $\sqrt{n}\,(\hat{\theta}_{n}- \theta^t)$ converge to the normal distribution $N(0, J_{\theta^t}^{-1})$ (asymptotic efficiency), where $J_{\theta}$ denotes the quantum Fisher information of the parameter $\th$, which turns out to be  16 for our model (\ref{eqn:model}).


%
%
\begin{figure}
\begin{center}
\includegraphics*[width=8.5cm]{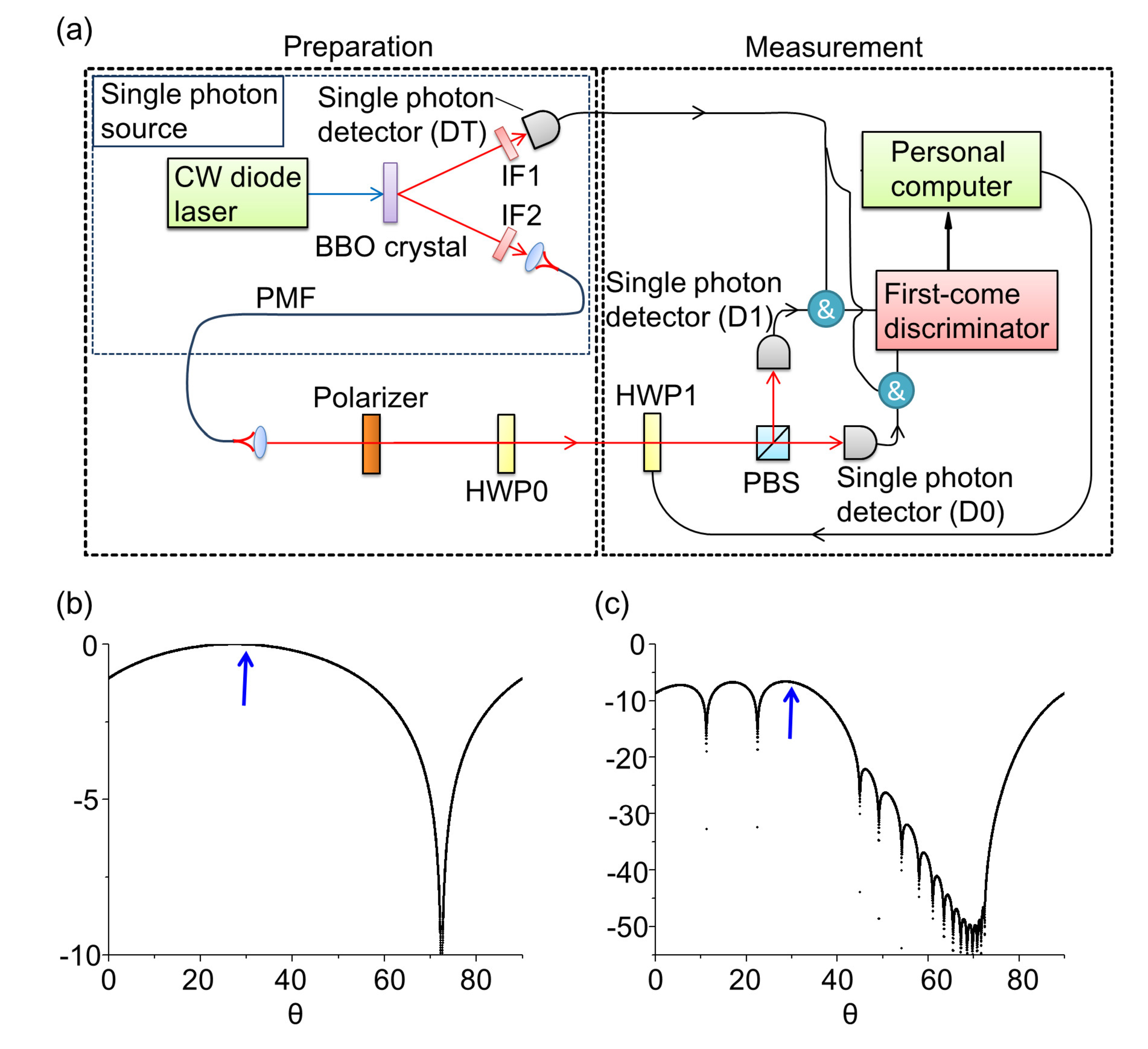}
\caption{(a) Schematic of the experimental setup. (b)(c) An example showing the update of a log-likelihood function. 
The second term $\log \bra{\psi(\theta)}M(x_n;\hat{\theta}_{n-1}) \ket{\psi(\theta)}$ in eq.~(\ref{eqn:elln}) is shown in panel (b), and the updated $l_{n}(\theta)$ is shown in panel (c). The blue arrows indicate the true value $\theta^t$.}
\end{center}
\end{figure}

The experimental setup is shown in Fig. 2(a).
Single photons at 780nm are generated from a heralded single photon source \cite{Hong-1986}, consisting of a CW diode pump laser (wavelength: 402 nm) and a 3 mm long BBO crystal (Type I). A pair of a signal photon (780 nm) and a trigger photon (830 nm) is created via spontaneous parametric down conversion. The detector (DT, SPCM-AQR, Perkin Elmer) after an interference filter (IF1, center wavelength 830nm) outputs an electric pulse (width 30ns) when it detects a trigger photon and the electric pulse heralds the generation of a signal photon, which is coupled to a polarization maintaining fiber (PMF) after an interference filter (IF2, center wavelength 780 nm, width 4 nm). The polarization of photons are then initialized to be horizontal using a polarizer (extinction ratio $10^{-5}$).
The target parameter $\theta^{t}$ was set using HWP0.
The polarization state of the photon was analyzed by HWP1 and a polarizing beam splitter (PBS).
After passing through the PBS, photons are guided to single photon detectors (D0 and D1, SPCM-AQR, Perkin Elmer) on each PBS output port. 
The outputs of single photon detectors are gated by the rise of the heralding signal and connected to the ``first-come discriminator,'' consisting of a home-made electric circuit.
When the discriminator receives the first signal from one of the detectors (D0 or D1) after the measurement for $(n-1)$th photon starts, the discriminator informs which detector has been clicked. 
The minimum pulse interval of 2.5ns can be discriminated. 
Note that the discriminator ignores the case when it receives the pulses from both the detectors within 2.5ns. The angle of HWP1 for measuring the $n$th photon is determined by calculating the discretized MLE $\hat\theta_n$, the maximizer of the log-likelihood function (\ref{eqn:elln}) chosen from among the 10000 points that divide the domain $[0, \pi/2)$ of the parameter $\th$ into equal parts (Figs. 2(b) and 2(c)).
When the change of HWP1 angle is completed, the measurement for the next ($n$th) photon will be started.  In a sequence of AQSE, the above mentioned procedure is carried out up to 300 input photons ($n$=300). For four different HWP0 angles $\theta=0, 30, 60$, and $78.3$ [deg], we repeated the sequence for 500 times ($r$=500).


Let us first observe the strong consistency for the sequence $\hat\th_n$ of MLEs for the parameter $\th$ of HWP0.
Fig. 3 (a) shows 500 trajectories of estimated HWP0 angle $\hat\th_n$ against the number $n$ of photons when the true value $\th^t$ of the parameter is set to be 60 degree. The curves correspond to independent runs of adaptive estimation. 
Evidently, each curve of $\hat\th_n$ approaches the true value $\th^t$, which is in accord with the mathematical result that $\hat\th_n\to\th^t$ almost surely as $n\to\infty$, even though the curves are dissimilar to each other reflecting the genuine statistical nature of quantum system. The convergence to the true value is clear in Fig. 3(b) where first 10 trajectories in Fig. 3(a) are superposed.


\begin{figure}
\begin{center}
\includegraphics*[width=8cm]{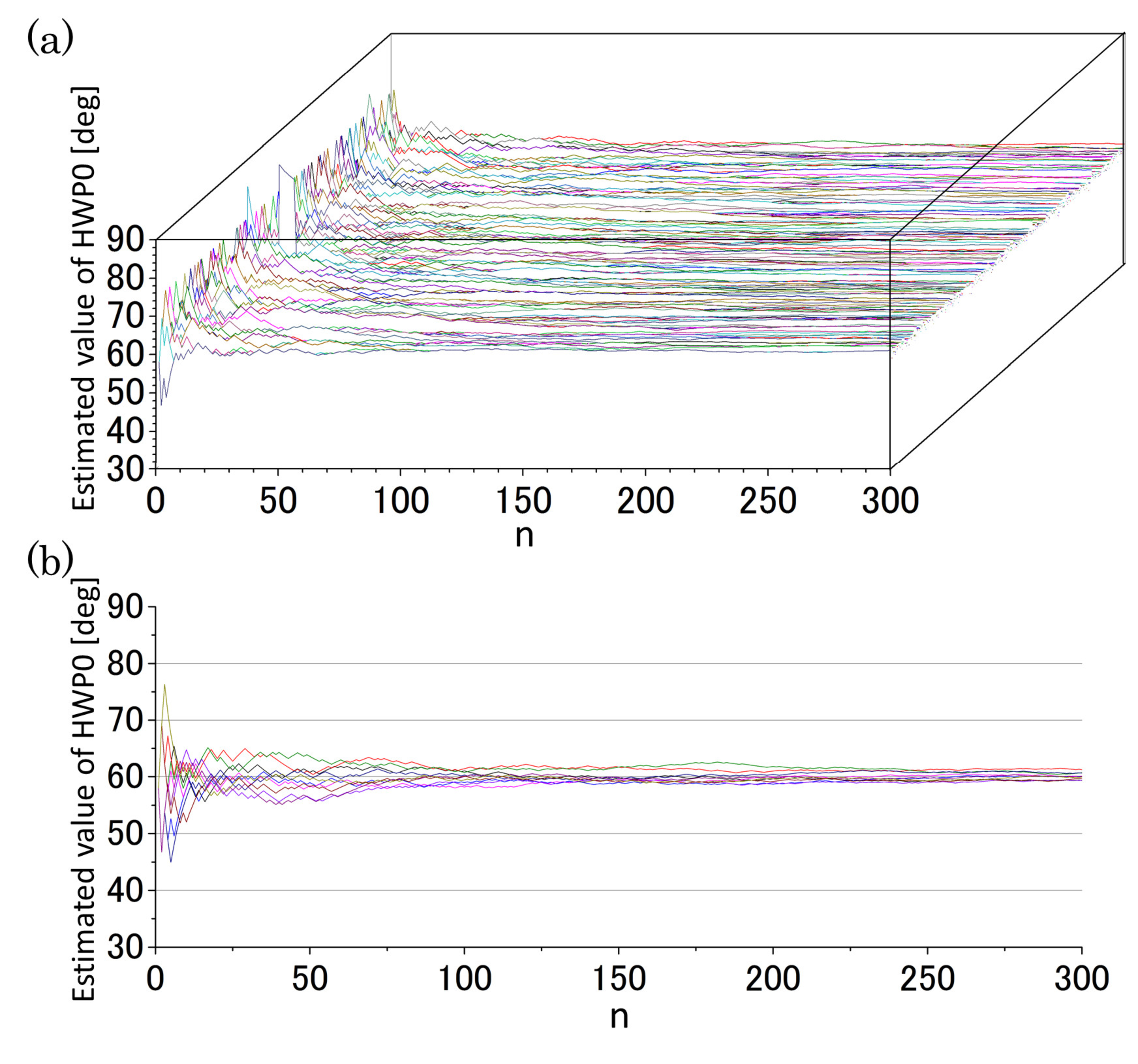}
\caption{(a) Trajectories of estimated HWP0 angles against the number $n$ of photons for $r=500$ repetitions is shown in a three dimensional plot. (b) The first 10 curves are superposed in a two dimensional graph.}
\end{center}
\end{figure}

%
%


We next test the hypothesis that the MLE $\hat\th_n$ follows a normal distribution for large $n$. More concretely, we will investigate if the random variable $\sqrt{n J_\th}\, (\hat\th_n-\overline\th)$ follows the standard normal distribution $N(0,1)$, i.e., 
$\sqrt{n J_\th}\, (\hat\th_n-\overline\th)\sim N(0,1),$ where $\overline\th$ is the sample average of MLEs $\hat\th_n$ over sufficiently many independent trials. 
A goodness of fit test \cite{statistics} was carried out as follows:

1) The real axis was divided into $23$ intervals (bins) $\{I_b\}_{b=0}^{22}$, 
where $I_1,\ldots, I_{21}$ are disjoint partitions of the interval $[-3.5, 3.5]$ of equal width, and $I_0=(-\infty, -3.5)$, $I_{22}=(3.5, +\infty)$. 
In reality, these bins were slightly shifted by $\d/10000$, where $\d:=\sqrt{n J_\th}\,\pi/20000$ is the scaled resolution of the estimator $\hat\th_n$, so that the data $\sqrt{n J_\th}\, (\hat\th_n-\overline\th)$ did not fall on the boundaries of the bins.

2) The test-statistic $X^2:=\sum_{b=0}^{22} \frac{(N_b-r\,p_b)^2}{r\,p_b}$ was calculated, where $N_b$ is the number of observed data which fell into $b$th bin, $p_b$ the theoretical probability of falling a datum into $b$th bin under the null hypothesis $N(0,1)$, and $r$ the number of repetitions of adaptive estimation procedure.

3) The test-statistic $X^2$ was analyzed using the chi-square distribution $\chi^2_{23-p}$ of degree $23-p$, where $p=2$ degrees of freedom ought to be subtracted because of the normalization and the use of sample average $\overline\th$.

\begin{figure}
\begin{center}
\includegraphics*[width=8cm]{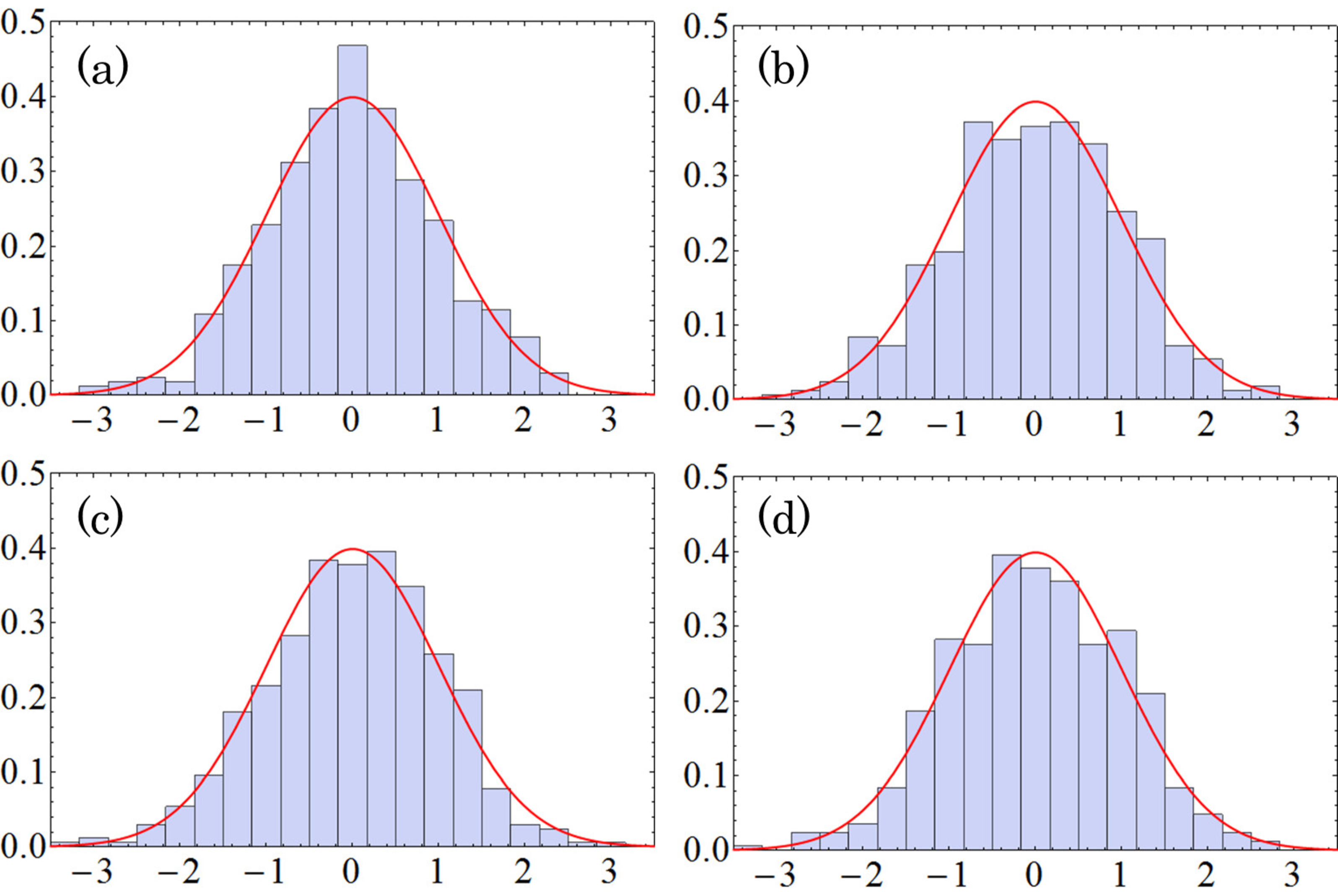}
\caption{Histogram of the observed data obtained by $r=500$ independent experiments of adaptive estimation scheme, each using $n=300$ photons. These histograms were taken for four different true values of (a) 0 [deg], (b) 30 [deg], (c) 60 [deg] and (d) 78.3 [deg].}
\end{center}
\end{figure}

Figure 4 shows the histogram of the observed data obtained by $r=500$ independent experiments of adaptive estimation scheme, each using $n=300$ photons. 
The true values $\th^t$ of the parameter $\th$ of HWP0 are set to be 0, 30, 60, and 78.3 degrees. 
The density function of the standard normal distribution $N(0,1)$ is also plotted as the solid curve. 
All the experimental data agree with the standard normal distribution.
To be precise, the values of the test statistic $X^2$ are (a) 16.8 (b) 15.7 (c) 12.8 (d) 16.2, and the null hypothesis is accepted with $10\%$ significance level in each case.


Having obtained the strong evidence that the distribution of the MLE has converged quite well to a normal distribution at $n=300$, we finally proceed to the estimation of confidence intervals \cite{statistics} for the mean $\m$ and variance $v$, assuming that $\sqrt{n}\, (\hat\th_n-\m) \sim N(0,v).$ The confidence intervals for $\m$ and $v$ are obtained by the standard procedure based on the statistical laws that $\sqrt{\frac{r}{\,\overline V\,}}\, (\overline\th-\m) \sim T_{r-1}$ and $\frac{r-1}{(v/n)}\, \overline V\sim \chi^2_{r-1}.$
Here $\overline V$ is the unbiased variance of MLEs $\hat\th_n$ over $r$ trials, and $T_{r-1}$ the $t$-distribution of degree $r-1$.


\begin{table}[t]
\begin{center}
{\bf \caption{Confidence intervals for the mean $\m$ and the variance $v$. CL means confidence level.}}
\begin{tabular}{lcclcclccl} \hline
 $\th^t$ [deg]  &\; \;\;& $\m$ [deg]  (90$\%$ CL) &\;\; \;& $v$  (90$\%$ CL) \\ \hline
   0.0  && -0.15 $\pm$ 0.06  && [0.054, 0.067] \\ \hline
 30.0  && 29.90 $\pm$ 0.06  && [0.055, 0.067] \\ \hline
 60.0  && 60.00 $\pm$ 0.06  && [0.056, 0.068] \\ \hline
 78.3  && 78.27 $\pm$ 0.06  && [0.055, 0.068] \\ \hline
\end{tabular}\\
\end{center}
\end{table}

Table 1 summarizes the results for $r=500$ with $90$\% confidence level.
Recall that the asymptotic efficiency asserts that $\m\simeq\th^t$ and $v\simeq J_{\th^t}^{-1}(=0.0625)$. 
Since the precision of the present experiment is about $\pm 0.2$ degree \cite{comment2}, we conclude that the estimated values of $\m$ and $v$ listed in Table I are in excellent agreement with the theoretical values.

%

It should be noted that the purpose of our AQSE is completely different from `adaptive measurements' proposed by Berry and Wiseman \cite{Berry2000}.
Their scheme was devised to estimate the phase difference between the two arms of an interferometer using a special $N$-photon two-mode state, approximating the canonical 
measurement proposed by Sanders and Milburn \cite{Sanders1995}, and is not applicable to general quantum state estimation problems. 
By contrast, our AQSE is a general-purpose estimation scheme applicable to any quantum statistical model using $n$ identical copies of an unknown state.
AQSE may also be used in verifying the achievability of the Cram\'er-Rao version of the Heisenberg limit $O(1/N^{2})$ \cite{ImaiFujiwara:2007} by applying the scheme to the $n$-i.i.d.~extension $\r_\th^{\otimes n}$ of an $N$-photon phase-shift model $\r_\th$ on $\H\simeq (\C^2)^{\otimes N}$. (See also \cite{FujiwaraImai:2008} for estimating a unitary channel under noise.) Incidentally, AQSE is based on the Cram\'er-Rao type point estimation theory and is free from the choice of {\em a priori} distribution which matters in Bayesian statistics such as adaptive Bayesian quantum tomography \cite{HusZarH:2012}.

In summary, we have verified both the strong consistency and asymptotic efficiency of AQSE by experimentally estimating the angle of linear polarization of photons. 
Since AQSE has been mathematically proven to outperform the conventional estimation scheme such as the state tomography \cite{Yamagata}, we plan to apply AQSE to multi-parameter cases and compare the performance with other protocols using fixed measurement basis \cite{Bogdanov-2010}. It will also be intriguing to apply AQSE to enhance the performance of quantum metrological experiments beating the standard quantum limit \cite{Giovannetti2004, Na-2007, Okamoto-2008, Xiang-2010}. 

We would like to thank Prof. Nagaoka for helpful discussion. This work was supported in part by Quantum Cybernetics project of JSPS, Grant-in-Aid from JSPS, JST-CREST project, FIRST Program of JSPS, Special Coordination Funds for Promoting Science and Technology, Research Foundation for Opto-Science and Technology, and the GCOE program. 


%
\end{document}